\documentclass[runningheads]{llncs}

\RequirePackage[bookmarksnumbered,unicode,hidelinks]{hyperref}
\RequirePackage{hyperxmp}

\PassOptionsToPackage{dvipsnames}{xcolor}
\usepackage[english]{babel}
\usepackage[utf8]{inputenc}
\usepackage{t1enc}
\usepackage{graphicx}
\usepackage[disable]{todonotes}
\usepackage{csquotes}
\usepackage{booktabs}
\usepackage{multirow}
\usepackage{fontawesome5}
\usepackage{paralist}
\usepackage{colortbl}
\usepackage{cleveref} % should be the last usepackage!

\newcommand{\todoi}[1]{\todo[inline]{#1}}

\newcommand{\review}{\textcolor{red}{REVIEW}}

\newcommand{\RQ}[1]{RQ#1}

\newcommand{\PSY}{\textsc{Psy}}
\newcommand{\SE}{\textsc{SE}}
\newcommand{\scopingreview}{\textsc{ScR}}
\newcommand{\GPT}{\textsc{GPT}}
\newcommand{\LLM}{\textsc{LLM}}
\newcommand{\SLR}{\textsc{SLR}}
\newcommand{\AI}{\textsc{AI}}

\newcommand{\cf}{\emph{cf.}}
\newcommand{\ie}{\emph{i.e.}}
\newcommand{\eg}{\emph{e.g.}}

\newcommand{\header}[1]{\textbf{#1}}

\newcommand{\answerblock}[1]{\begin{center}\noindent\fbox{\parbox{\textwidth}{#1}}\end{center}}

\begin{document}
	\title{Show Your Title! A Scoping Review on Verbalization in Software Engineering with \LLM-Assisted Screening}
	\titlerunning{Show Your Title! A Scoping Review on Verbalization in \SE{} with \LLM}
	
	\author{Gergő Balogh\thanks{Corresponding author.}\orcidID{0000-0002-6781-5453} \and Dávid Kószó\orcidID{0000-0003-0617-8160} \and Homayoun {Safarpour Motealegh Mahalegi}\orcidID{0000-0002-3206-5357} \and László Tóth\orcidID{0000-0002-5797-8670} \and Bence Szakács\orcidID{0009-0009-2712-0487} \and Áron Búcsú\orcidID{0009-0001-5736-3748}}
	\authorrunning{Gergő Balogh et al.}

	\institute{Department of Software Engineering, University of Szeged,\\Árpád tér 2., 6720 Szeged, Hungary \\\email{\{geryxyz,koszod,safarpour,premissa,szakacs,baron\}@inf.u-szeged.hu}}

	\maketitle
	
	% \sloppy
		
	\begin{abstract}
		Understanding how software developers think, make decisions, and behave remains a key challenge in software engineering (\SE). Verbalization techniques (methods that capture spoken or written thought processes) offer a lightweight and accessible way to study these cognitive aspects. This paper presents a scoping review of research at the intersection of \SE{} and psychology (\PSY), focusing on the use of verbal data. To make large-scale interdisciplinary reviews feasible, we employed a large language model (LLM)-assisted screening pipeline using GPT to assess the relevance of over 9,000 papers based solely on titles. We addressed two questions: what themes emerge from verbalization-related work in \SE, and how effective are LLMs in supporting interdisciplinary review processes? We validated GPT's outputs against human reviewers and found high consistency, with a 13\% disagreement rate. Prominent themes mainly were tied to the craft of \SE, while more human-centered topics were underrepresented. The data also suggests that \SE{} frequently draws on \PSY{} methods, whereas the reverse is rare.

	\keywords{%
		scoping review \and
		verbalization \and
		software engineering \and 
		psychology \and 
		cognitive aspects \and 
		knowledge representation \and 
		meta-research \and 
		software development practices
	}
    \end{abstract}

	\section{Introduction}
	\todoi{\review{} It is unclear how/why verbalization is any different for SE than it would be in other (humanity) disciplines. One way in which it could be would be to link verbalization to cognitive processes/theories. But the paper does not go into any depth in that direction.}
	
	\todoi{\review{}  some references are sloppy.}
	
	\todoi{\review{} The motivation behind this systematic literature review is poor and it should be motivated with scientific publications. For me, there should be a concrete example as to why this systematic literature review.}

	Software engineering (\SE) research often relies on implicit assumptions about how developers interact with tools, documentation, and processes. However, upon closer examination, these assumptions may not always align with real-world behavior. Our research has repeatedly encountered surprising moments where experimental results contradicted our underlying beliefs, sometimes even unconscious ones. These moments were more frequent when studying human factors, where software engineers' cognitive processes and decision-making patterns do not always conform to our expectations.
	A recurring pattern emerges: We assume developers think and act in a certain way, designing tools and methodologies accordingly. Sometimes, we are correct, but other times, we are not. Why do they prefer one approach over another? Why do they disregard seemingly efficient tools in favor of alternative methods? These are fundamental questions for \SE{} research, yet they are often left unexplored, and only a few studies take into account the cognitive processes in their investigation (\cf{} \cite{cognitive-biases-in-se,feelings-in-se}).

	For instance, fault localization research often assumes developers investigate ranked lists of suspicious statements in order. In reality, do they systematically follow these suggestions, or do they jump around based on intuition and experience \cite{using-contextual-knowledge-in-interactive-fault-localization}? Is this due to habit, ease of use, or deeper issues related to cognitive load? Preliminary observations reveal that expert debuggers rarely follow a strictly linear, sequential approach when inspecting suspicious code lines. Instead, they iterate through hypothesis testing, jump between code segments, and prioritize key areas based on intuition \cite{Park.2024}. This method may be due to cognitive load management \cite{Gao.2023}, where relying on experience allows them to quickly eliminate incorrect paths rather than following a strict, line-by-line debugging process \cite{Bednarik.2012}.

	We propose leveraging verbalization techniques: methods that capture developers' thoughts in spoken or written form during work. These techniques allow researchers to make observations without specialized equipment like eye-trackers or medical imaging devices, hence could be more accessible. Our goal is to bridge the knowledge gap between humanities experts and \SE{} researchers to enhance understanding of developers' cognitive processes, behaviors, and decision-making patterns. We assume that an in-depth analysis will yield more effective \SE{} theories, widely adopted tools, and a more productive industrial environment in the long run. Our first step toward this goal is to map the intersection between the two fields: \SE{} and psychology (\PSY). This paper reports on our scoping review (\scopingreview{}), explained in \cite{scoping-studies-towards-a-methodological-framework}, about involving verbalization or acquiring verbal data from participants in the context of \SE.
	
	We define verbalization and verbal data based on entries from \cite{OED,MerriamWebster}. \emph{Verbal data} refers to information expressed through spoken or written language, including speech and text used for analysis. \emph{Verbalization} is the process of expressing thoughts or ideas in words, whether spoken or written.

	\subsection{Research Questions} \label{sec:research-questions}
	
	Now, we briefly overview our contribution. In this paper, we seek to answer the following two research questions. 

	\begin{enumerate}
		\item[\RQ1:] \emph{What prominent themes emerge from the intersection of \SE{} and \PSY{} literature related to verbalization, when viewed from a high-level scoping perspective?}

		\item[\RQ2:] \emph{How well can large language models (\LLM s), especially \GPT, perform and support consistent, title-based relevance screening and thematic exploration in broad interdisciplinary \scopingreview{}s?}
	\end{enumerate}
	
	Our professional experiences let us assume, that the intersection of \SE{} and \PSY{} (especially in the context of verbalization-based techniques) is complex and has not been reviewed comprehensively before. This \scopingreview{} serves a dual purpose: to map the existing interdisciplinary literature at the intersection of \SE{} and \PSY{} related to verbalization (\cf{} \RQ1), and to explore the feasibility of novel methodological approaches, such as \LLM-assisted screening and title-based relevance assessment (\cf{} \RQ2). This combined focus enables us to both innovate in review techniques and provide an initial landscape of the field. For \RQ1 and \RQ2, the in-depth explanation can be found in \Cref{sec:literature-review} and \Cref{sec:validation}, respectively.
	
	Please note that our research does not aim to establish a benchmark for \GPT{} in interdisciplinary \SE{} research systematic reviews. But all preliminary and raw data, along with the scripts used, is made available to facilitate further research \cite{online-appendix}.

	\section{Scoping Literature Review Process}
	\label{sec:literature-review}

	\todoi{Related work:

		- Process: \cite{slr-process-kitchenham,slr-process-ampat,slr-process-sjoberg}

		- Examples: \cite{slr-agile-software-development,ex-slr-agile-software-development, slr-gamification-in-se,slr-large-scale-agiel-sd}
	}

	\begin{figure}[h]
		\centering
		\includegraphics[width=\textwidth]{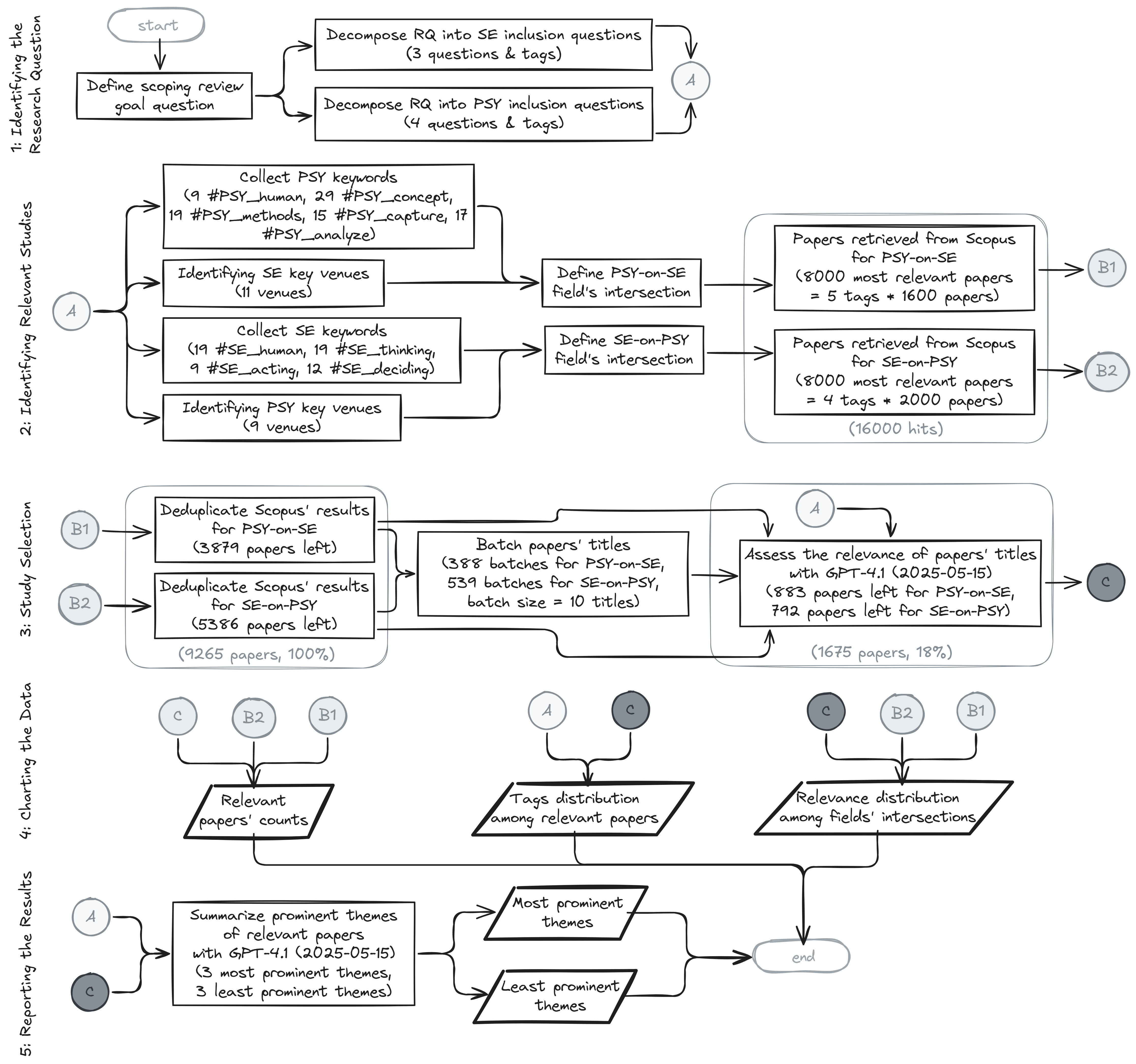}
		\caption{Flowchart of our \scopingreview{} process with the five stages}
		\label{fig:systematic-review-process}
	\end{figure}

	In this section, we explain the details of our \scopingreview{} process depicted in \Cref{fig:systematic-review-process}. We adopt the widely used framework \cite{scoping-studies-towards-a-methodological-framework}, which defines scoping studies as a way to map the extent, range, and nature of research activity in a field. Their five-stage model guides our process, adapted for the interdisciplinary and computational nature of our review. Recall that our goal is to bridge \PSY{} and \SE{} research, and then spread out from that common ground to the closest parts of the two fields in the upcoming systematic literature review (\SLR).

	Note that a \scopingreview{} differs from a \SLR{} mainly in purpose and scope: a \SLR{} aims to answer a focused research question by critically appraising and synthesizing high-quality evidence. Guidelines for conducting a \SLR{} in the field of \SE{} can be found in \cite{kitchenham-base,slr-process-brereton,ex-slr-agile-software-development,slr-process-kitchenham-2009,slr-process-kitchenham}. For examples of a \SLR{} in the field of \SE, we refer to \cite{slr-agile-software-development,slr-gamification-in-se,slr-large-scale-agiel-sd}. Observe that, despite their differing objectives, \scopingreview s and \SLR s frequently employ similar methodological approaches.

	\subsection{Identifying the Research Question} \label{sect:identify-research-questions}
	
	This phase in the \scopingreview{} process is foundational, as it establishes the scope and purpose of the entire review and guides the discovery process. In our case, the central research question is \RQ1 (\cf{} Stage 1 of \Cref{fig:systematic-review-process}). By \enquote{prominent themes}, we mean broad, recurring or conceptual linkages that surface across the intersection of \SE{} and \PSY{} literature, highlighting how verbalization is positioned. This question is intentionally scoped as a bird's-eye view to create a raw map of topics and potential research directions to pursue or explore with further in-depth \SLR. \RQ1 defines the goal of the \scopingreview{}, while \RQ2 focuses on evaluating the validity of our methodology. As such, \RQ2 is only mentioned here and will be detailed in the upcoming sections on validation (\cf{} \Cref{sec:validation}).

	To operationalize \RQ1, we developed seven inclusion questions (three from \SE{} and four from \PSY), each tagged for clarity and systematic application. Now, we present each inclusion question with its tag, which will be used later in the process to refer to the inclusion question.
	
	\texttt{\#\SE{}\_thinking}: \emph{Does the paper have applications in human cognition during software development or \SE{} experts' cognition?}

	\texttt{\#\SE{}\_acting}: \emph{Is the context related to human behavior during software development or \SE{} experts' behavior?}

	\texttt{\#\SE{}\_deciding}: \emph{Is the context related to human decision-making during software development or \SE{} experts' decision-making?}
	
	\texttt{\#\PSY\_concept}: \emph{Does the paper involve turning thoughts into words or collecting spoken/written words?} 

	\texttt{\#\PSY\_methods}: \emph{Does the paper discuss methods for capturing participants' thoughts, actions, and decisions?} 

	\texttt{\#\PSY\_capture}: \emph{Does the research use participants' words to record their thoughts, actions, and decisions?} 

	\texttt{\#\PSY\_analyze}: \emph{Does the paper aim to understand the participants' thoughts, actions, and decisions using their words?}
	
	For brevity, their full, detailed versions are included in the online appendix \cite{online-appendix}. We applied all questions to both \SE{} and \PSY{} literature to ensure a comprehensive mapping of this intersection. The shared prefixes on their tags only denote their origin, not their intended usage. These inclusion questions served as practical checkpoints, shaping the collection, relevance assessment, and theme mapping throughout our review. They help ensure consistency and thoroughness at most of the steps.

	\subsection{Identifying Relevant Studies} \label{sect:identify-relevant-studies}

	To capture the intersection of the two fields we inspected the papers matched to \PSY-related keywords but published on any of the \SE{} venues (\PSY-on-\SE). Similarly, we define \SE-on-\PSY{} as the papers matched to \SE-related keywords but published on any of the \PSY{} venues.

	By considering the seven inclusion questions given in \Cref{sect:identify-research-questions}, for each inclusive question, we collected related keywords. During the collection, we were striving to answer the question: \enquote{If this keyword appears in a paper, which inclusion question would I most confidently answer \enquote{yes} to because of it?}
	We mention that while we were collecting the keywords, we found that some keywords may belong to more than one inclusion question within the domain. To avoid confusion, we included two additional tags: \texttt{\#\SE\_human} and \texttt{\#\PSY\_human}. The former is used for keywords related to human factors on the \SE{} side, while the latter is used for keywords related to \PSY{} human factors only. (We note that these two additional tags are not inclusion questions but rather a way to categorize the keywords.)
	
	Some keywords contain a \texttt{*} character, which has a special meaning for the Scopus Search API\footnote{\url{https://dev.elsevier.com/documentation/SCOPUSSearchAPI.wadl}, accessed: 2025-05-23.} (utilized for collecting papers). In that context, \texttt{*} can represent an arbitrarily long sequence of characters, \eg{}, the keyword \texttt{*verbal} represents both the keywords \texttt{verbal} and \texttt{intraverbal}.) The distribution of the collected keywords per field is shown in Stage~2 of \Cref{fig:systematic-review-process}.
	For the complete list of keywords used, consult the online appendix \cite{online-appendix}.

	In parallel, we collected venues that may publish papers belonging to \SE{} or \PSY{} fields. Initially, we used our experience to select relevant \SE{} venues, preferably with CORE A\footnote{\url{https://portal.core.edu.au/conf-ranks/}, accessed: 2025-05-23.} rankings. In this process, we might include other venues in the list even if they do not have rank A but are deemed prominent by the researchers involved in this paper. As a result, in the case of \SE, we collected eleven venues (\cf{} Stage~2 of \Cref{fig:systematic-review-process}). 
	Unlike in \SE{}, we could not find any universal ranking of \PSY{} venues. Hence, we selected several prominent groups and associations (the \PSY{} counterparts of \textsc{acm} and \textsc{ieee}) and used their list of supported venues to assemble our list of \PSY{} key venues. We included all the flagship journals and conferences of these associations, along with additional venues that might be relevant to verbalization-based research methods, striving for a very inclusive approach to ensure we did not overlook any potentially important sources. We collected nine venues (\cf{} Stage~2 of \Cref{fig:systematic-review-process}). We note that the selected venues are not exhaustive, but we assume that they are representative of the fields. The selected venues per field are available in the online appendix \cite{online-appendix}.

	Then, we used the Scopus Search API to collect papers.
	We note that several parameters can be set here to customize the search query. For us, only two were of interest: \texttt{query} and \texttt{sort}. To define \PSY-on-\SE{} field's intersection (\cf{} Stage~2 of \Cref{fig:systematic-review-process}), we constructed each search query by setting the parameter \texttt{query} as follows. For each tag that starts with \texttt{\#\PSY{}\_}, we took the collected keywords related to that tag, joined them by the usual logical operator \texttt{OR}, and the collected venues of the field \SE{} joined by the usual logical operator \texttt{OR}, and connected them. We utilized the \texttt{TITLE-ABS-KEY} and the \texttt{SRCTITLE} Scopus field codes for keywords and venues, respectively. To find more relevant papers, we also set the parameter \texttt{sort} to \texttt{relevancy}\footnote{\url{https://service.elsevier.com/app/answers/detail/a_id/14182/supporthub/scopus/kw/relevance/}, accessed: 2025-05-23.}. Vice versa, to define \SE-on-\PSY{} field's intersection (\cf{} Stage~2 of \Cref{fig:systematic-review-process}), we repeated the same process, but now we took the collected keywords related to each tag starting with \texttt{\#\SE{}\_} and the collected venues of the field \PSY{}.

	\subsection{Study Selection} \label{sec:study-selection}

	In this subsection, we explain how we selected relevant papers from the collected ones (\cf{} Stage~3 of \Cref{fig:systematic-review-process}). A paper can be found by the Scopus Search API multiple times as it can be relevant to multiple keywords, and thus, it is natural to deduplicate the collected papers. Since each collected paper has a unique identifier (the \texttt{eid} provided by Scopus), we used that for deduplication.
	
	Then, as we mentioned in \Cref{sec:research-questions}, we propose an  \LLM-assisted screening and title-based relevance assessment. For this, we utilized GPT-4.1 (2025-05-15) (referred to as: \GPT) because it is a state-of-the-art and general purpose model. To optimize our request for \GPT{} tokens, we decided to create batches of $10$ titles. We manually checked with some examples that \GPT{} is able to process that many titles in one batch. Our prompt for \GPT{} was designed to be concise and focused, besides the $10$ titles of a batch, it also contains the seven inclusion questions given in Section~\ref{sect:identify-research-questions}. We utilized the well-known Persona pattern \cite{persona-pattern} to instruct \GPT{} to act as \enquote{an expert in humanities and software engineering}.
	
	For each collected paper in the deduplicated set, the task of the \GPT{} was to decide whether that paper could be relevant or not considering only its title and the seven inclusion questions (both \SE{} and \PSY{} ones, since our interdisciplinary research aims to capture the intersection). Hence, for each paper, the simplified output of the \GPT{} was a JSON file with the following three  fields: \texttt{relevance}, \texttt{justification}, and \texttt{tags}. 
	
	The value of the field \texttt{relevance} is binary: \texttt{relevant} or \texttt{not relevant}. The value of the field \texttt{justification} consists of 1-2 sentences about why \GPT{} thought that paper was relevant or not. We mention that the value of that field is also crucial for later steps (\cf{} \Cref{sec:validation,sec:reporting-results}).
	Finally, the value of the field \texttt{tags} is a possibly empty array of the tags of those inclusion questions (\cf{} \Cref{sect:identify-research-questions}), for which \GPT{} answered \enquote{yes}. (Observe that if a paper is not relevant, then that array is empty.)

	Since \GPT{} can make mistakes, we repeated this phase three times to see how consistent \GPT{} is with itself, and then aggregated the results by majority voting. For further results, please, consult the online appendix \cite{online-appendix}.
	\todoi{@Gergo: please, add a reference to the aggregated results, too.}
	
	\subsection{Charting the Data}
	\label{sec:charting-data}

	Following \Cref{fig:systematic-review-process}, this subsection presents the statistics of the collected papers (\cf{} Stages~2 and 3). Upon completion of Stage~2, we collected $8,000$ \SE-on-\PSY{} papers, and $8,000$ \PSY-on-\SE{}, \ie{}, $16,000$ papers in total. More precisely, $2,000$ papers were collected for each keywords set (grouped by a tag) from \SE{}, and $1,600$ papers for each keywords set from \PSY{} (\cf{} \Cref{sect:identify-research-questions,sect:identify-relevant-studies}). In both cases, we found that the first $8,000$ search results contain most of the relevant papers by trial and error; the rest can be safely ignored.
	Recall that our approach heavily relies on the search engine relevance calculated on Scopus (\cf{} \Cref{sect:identify-relevant-studies}).

	\begin{figure}
		\todoi{@Gergo: please, regenerate this figure with hypens, cf. PSY-on-SE and SE-on-PSY}
		\centering
		\includegraphics[width=0.8\textwidth]{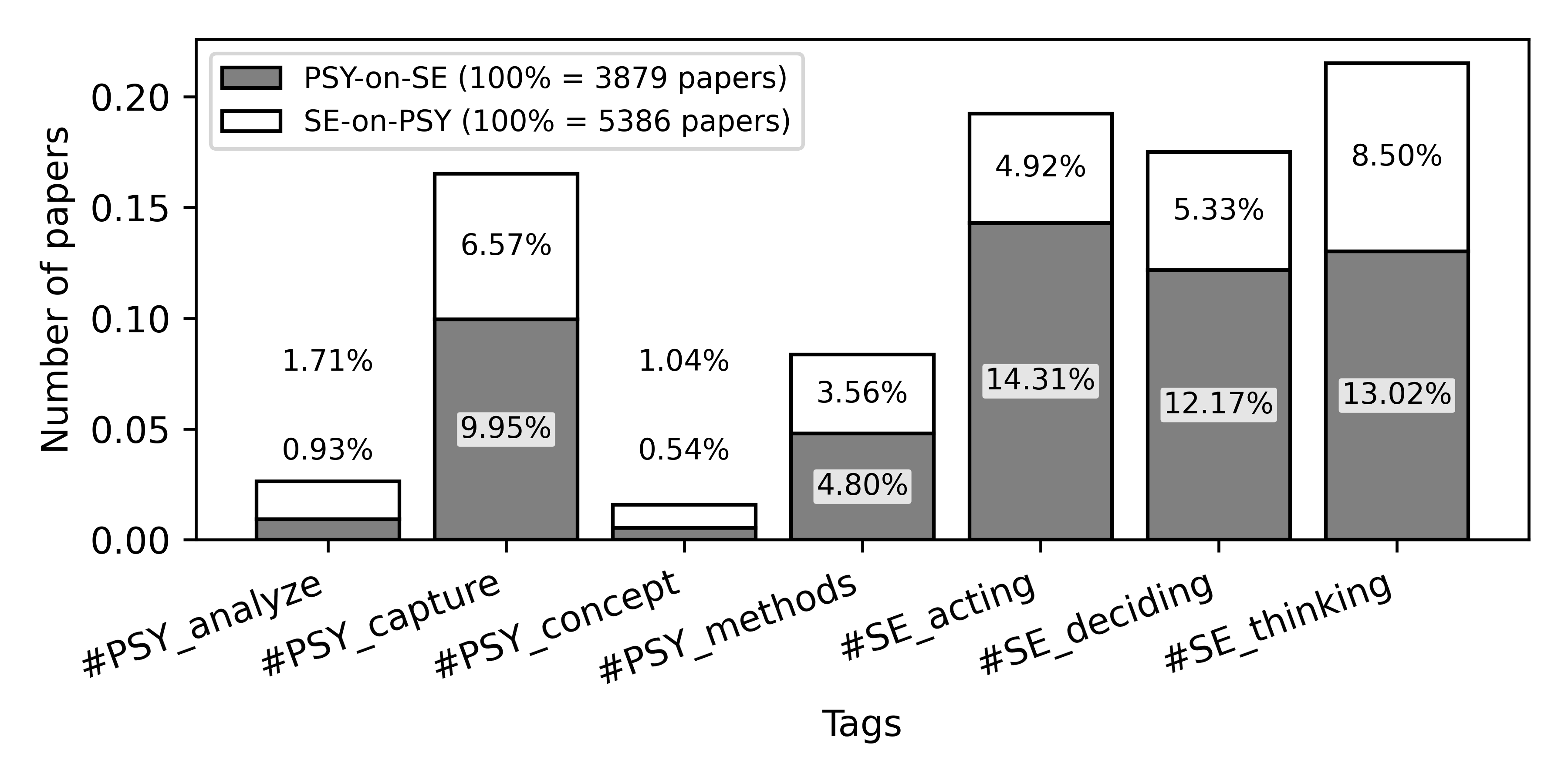}
		\caption{Normalized distribution of the inclusion question tags of the relevant papers}
		\label{fig:tags-distribution}
	\end{figure}

	Now, we continue with the statistics of Stage~3 of \Cref{fig:systematic-review-process}. After the deduplication process, $5,386$ \SE-on-\PSY{} and $3,879$ \PSY-on-\SE{}, \ie{}, in total $9,265$, papers remained. Then we created batches of $10$ titles (\cf{} \Cref{sec:study-selection}): we had $539$ batches for \SE{} and $388$ batches for \PSY{}.

	Finally, we give the statistics promised for Stage~4 of \Cref{fig:systematic-review-process}. After \GPT{} processed the batches in three rounds, $1,675$ papers were marked as relevant (\cf{} Stage~3 of \Cref{fig:systematic-review-process}).
	\Cref{fig:tags-distribution} shows the distribution of the inclusion question's tags of the relevant papers. Since we found more papers in the \SE-on-\PSY{} intersection we normalized the count of tags by the number of papers retrieved from that particular intersection. Recall that a paper could be marked with more than one tags, since it could be relevant based on multiple inclusion questions. For further details on the distribution, please, consult the online appendix \cite{online-appendix}.
    Eventually, we discuss the relevance distribution among the intersections of the two fields. Out of $5,386$ papers in the \SE-on-\PSY{} intersection, $792$ ($14.7\%$) were marked as relevant by \GPT{}. Similarly, out of $3,879$ papers in the \PSY-on-\SE{} intersection, $883$ ($22.76\%$) were marked as relevant. \todoi{what does numbers mean?\\
	@Gergo: It is explained in the second paragraph of the next subsection}

	\subsection{Reporting the Results}
	\label{sec:reporting-results}
	
	We summarize the results of our \scopingreview{} according to Stage~5 of \Cref{fig:systematic-review-process}. We discuss the distribution of inclusion questions' tags presented in \Cref{fig:tags-distribution}. Observe that it reveals a notable asymmetry in the engagement between the fields of \PSY{} and \SE{}. Specifically, \PSY-related keywords appear more frequently in \SE{} venues than \SE-related keywords do in \PSY{} venues (ratio of grey versus white bars in \Cref{fig:tags-distribution}). This suggests that \SE{} research is increasingly incorporating \PSY{} constructs. These findings suggest that while \SE{} scholars actively draw on \PSY{} theories and methods, the inverse is less common, revealing an opportunity to foster deeper bidirectional collaboration and increase the visibility of \SE{} insights within \PSY{} research. This asymmetry is also justified by the fact that we found more relevant papers in the \PSY-on-\SE{} intersection than vice versa, as discussed at the end of \Cref{sec:charting-data}. 

	Eventually, we identify research gaps and directions to follow. Here, \GPT{} has played a crucial role again as we instructed it to group the relevant papers based on their \texttt{justification} field, give a meaningful theme name for each group with an explanation and a reason for prominence, and select the top $3$ and bottom $3$ themes for further evaluation. We repeated this process for each of the three runs of \GPT{}, and we collected the results in the online appendix \cite{online-appendix}.

	As a final point, we answer \RQ1, which asks about the prominent themes emerging from the intersection of \SE{} and \PSY{} literature related to verbalization.
	For this, we manually aggregated the top $3$ and bottom $3$ themes from the three runs of \GPT, and selected the most and the least prominent ones.

	The most prominent themes are the following.
	\emph{Expert Cognition, Decision-Making, and Behavior} explores the cognitive processes, decision-making strategies, and behaviors of \SE{} experts, highlighting how expertise and cognitive biases shape their strategic problem-solving approaches.
	\emph{Use of Psychological or Qualitative Methods} examines the application of psychological theories and qualitative research methods in \SE{}, highlighting the importance of understanding human behavior and cognition in the development process.
	\emph{Teamwork, Collaboration, and Group Cognition} explores the dynamics of teamwork and collaboration in \SE{}, focusing on how group cognition and social interactions influence project outcomes.
	
	The least prominent themes are as follows.
	\emph{Indirect or Implicit Human Factors} addresses the less visible aspects of human factors in \SE{}, such as implicit biases and unspoken assumptions that influence decision-making and behavior.
	\emph{Relevance by Association to Diversity, Inclusion, or Demographics} explores how diversity, inclusion, and demographic factors are associated with \SE{} practices, highlighting the need for a more inclusive approach to understanding human factors.
	\emph{Historical or Foundational Context} examines the historical and foundational aspects of \SE{}, focusing on how past practices and theories shape current understanding and approaches to human factors in the field. 

	Our \scopingreview{} revealed some intriguing, although not always generalizable, patterns between the most and least prominent themes in \SE{} research. Specifically, the most prominent themes are deeply rooted in the art and practice of \SE{}, treating human involvement as a supporting element for technical processes. In contrast, the least prominent themes are driven primarily by human-centered research interests, only later intersecting with \SE. These differences may partly reflect the innate distinctions between \PSY{} and \SE{} as disciplines, where \SE{}'s practical focus can sometimes lead to using \PSY{} tools and concepts without a solid theoretical foundation. This raises our concern that rigorous research, especially in subfields where human factors play a central role, must not allow methodologies to be employed without fully understanding their principles and implications. Overall, while human factors are recognized across these themes, the way they are integrated into \SE{} research suggests both opportunities and caution for future studies.

	\answerblock{
	\header{In answering \RQ1,} our \scopingreview{} identified a contrast between the prominent themes rooted in the craft of \SE{} (such as expert cognition, decision-making, and teamwork) and the more human-centered, least prominent themes like indirect human factors, diversity, and historical contexts. This suggests promising yet underexplored research gaps worth future exploration. We also observed an imbalance in the number of inclusion questions answered across the \SE{} and \PSY{} domains, although the reasons behind this remain unclear. We suppose that \SE{} often transfers actionable psychological knowledge into practice, while \PSY{} does not typically investigate \SE-related human factors, though these are assumptions that should be verified in future studies.
	}
	
	\section{Validation of Our Approach}
	\label{sec:validation}
	
	Recall that our \RQ2 summarizes the overarching goal of validating our \scopingreview{} methodology with \LLM{} (\GPT). This question aims to assess both the validity and evaluate the practical usefulness of \LLM s, especially \GPT, in supporting relevance screening and thematic exploration within broad, interdisciplinary \scopingreview{}s. Given the vastness of the field, we are particularly interested in how \LLM s can help manage resource constraints while still providing reliable and meaningful guidance in the review process. The goal is to understand whether \LLM s offer a feasible balance between efficiency and accuracy.
	
	\begin{figure}[h]
		\todoi{squeeze}
		\centering
		\includegraphics[width=\textwidth]{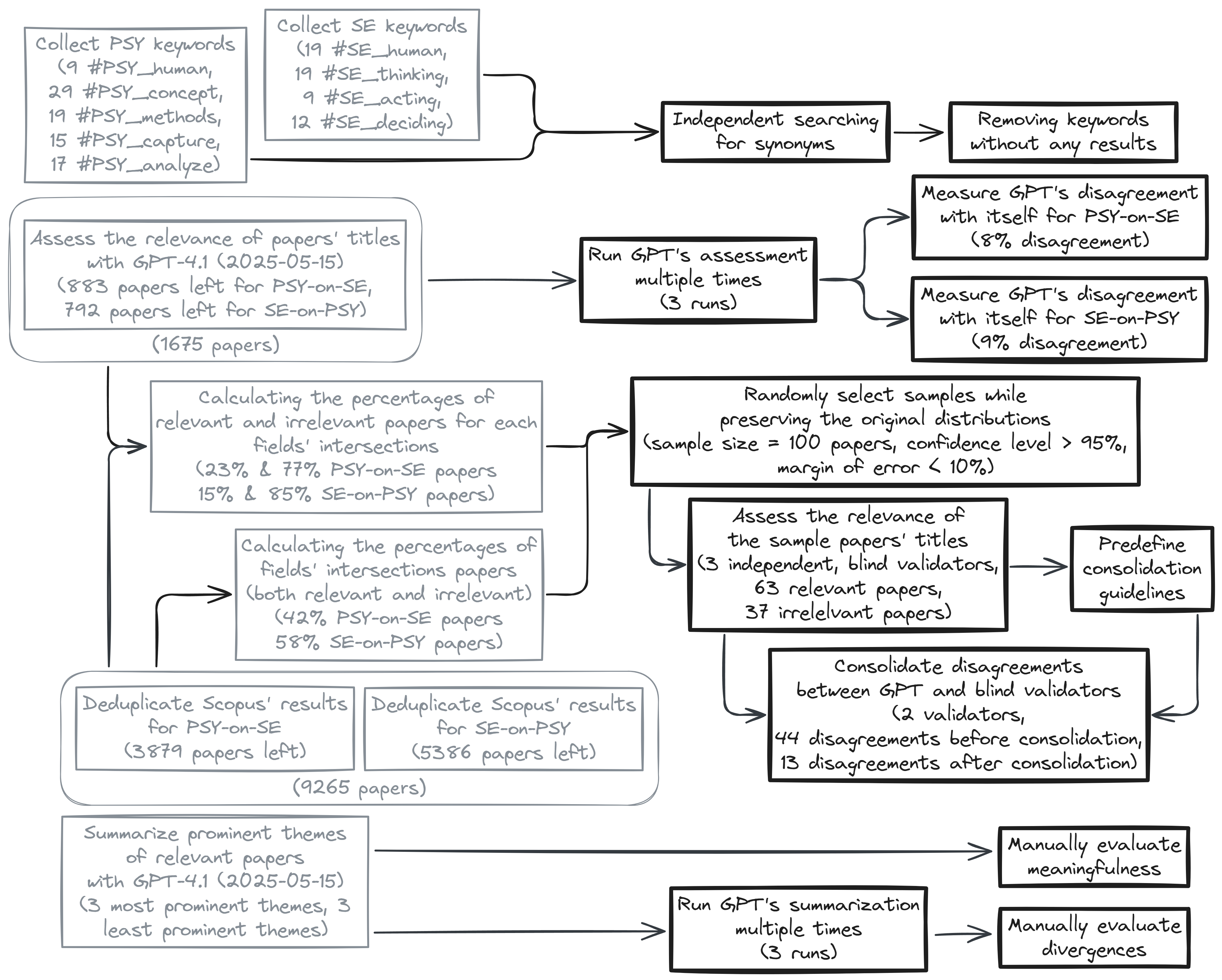}
		\caption{Flowchart of our \scopingreview{}'s validation process (grey = review process step, bold black = validation step)}
		\label{fig:validation-process}
	\end{figure}
	
	As noted in \Cref{sec:literature-review}, a search using only relevant keywords and venues can return thousands of results, most of which are irrelevant. Researchers face a practical challenge: reducing either the number of papers to screen or the amount of information each one requires for assessment, or both. Our \scopingreview{} encountered a dilemma, heuristics are needed to filter results, but developing them requires high-level domain knowledge, which is also the goal of our review. To address this, we propose automating part of the screening process using an \AI{} reviewer. To further reduce effort and resource use (\eg, token cost), we limit the input to paper titles. Our validation is twofold: we assess whether \GPT{} can generate useful insights from minimal input and whether it can do so consistently under these constraints.
	
	\subsection{Assess Usefulness}
	\label{sec:asses-usefulness}
	
	We validated the usefulness of our approach in two ways. Firstly, we manually evaluated the sample set of titles to ensure \GPT{} relevance assessment is aligned with the human reviewers (\cf{} \Cref{sec:study-selection}). Secondly, we utilize our expertise to judge if our \scopingreview{} process yields any meaningful prominent themes (\cf{} \Cref{sec:reporting-results}). Since this present research is a precursor to an in-depth \SLR{} study, we considered a theme meaningful if it is sensible to follow in the upcoming research.
	
	We selected a sample of 100 papers to reflect the overall distribution of intersections in our dataset, ensuring representation across relevant and irrelevant papers. Specifically, 42\% of the sample came from \PSY-on-\SE{} papers, while 58\% were \SE-on-\PSY{} papers. Within these, 23\% of \PSY-on-\SE{} papers were relevant, compared to 77\% irrelevant, and \SE-on-\PSY{} papers had 20\% relevant and 80\% irrelevant. This sampling  preserved the original proportions of each intersection and maintained a confidence level above 95\% with a margin of error of less than 10\%. \todoi{stat sig.}
	
	Three independent, blind validators manually assessed the titles in this sample, classifying 63 papers as relevant and 37 as irrelevant. These validators followed the same instructions as \GPT, although they did not process papers in batches like \GPT{} did since the batching was solely to manage \GPT's token usage. Before any consolidation, we noted 44 disagreements between \GPT's assessments and those of the human reviewers. To address these, two \SE{} experts examined the disagreements and discussed them in the context of predefined consolidation guidelines. These guidelines prioritized strictly \SE-related content and, in cases of uncertainty, retained papers for potential exclusion during a later full systematic review.
	
	Most disagreements arose because \GPT{} was more conservative in its assessments, often rejecting papers the human reviewers accepted—particularly those focusing more on the user side rather than the developer perspective. After consolidation, only 13 disagreements remained, amounting to a 13\% disagreement rate. We considered this level of disagreement acceptable, given the complexity of assessing relevance across intersecting domains. This outcome underscores the usefulness of \GPT's assistance in large-scale \scopingreview{}, especially when working with broad, interdisciplinary datasets.

	We also considered the credibility and believability of the most and least prominent themes identified among the relevant papers. We agreed that these themes, as detailed in \Cref{sec:reporting-results}, make sense in the broader \SE{} and \PSY{} landscape and align well with the practical challenges and research interests in the field. Furthermore, we believe these themes could serve as meaningful guides and even form the foundation for developing heuristics for our upcoming \SLR. Overall, they provide a promising starting point for shaping more targeted and structured exploration in future studies.
	
	\subsection{Examine Consistency}
	
	Our perspective is that for any scientific approach to be reliable and valuable within the \SE{} community, it must be consistent. To assess this, we analyzed the consistency of both \GPT{} and human reviewers. We compared the answers and opinions of each to their peers and each other.
	
	For \GPT's internal consistency, we ran the same queries three times in both the relevance assessment and theme summarization stages, using default vendor settings. Divergences were counted when the three outputs differed in relevance (\eg, two relevant votes, one irrelevant). For theme summarization, we subjectively reviewed phrasing differences to determine if they still represented the same concept.
	
	For human reviewers, we compared their relevance votes on the same paper titles. Even though the sample size (three reviewers) is limited, it provides insight into human variability. According to the manual validation dataset, perfect agreement on relevance was achieved in 60\% of cases, while 40\% of cases showed partial disagreements (\Cref{sec:asses-usefulness} provides sample selection details). \GPT's achieved perfect agreement (with itself) on relevance in 91\%-92\% of cases, with 9\%-8\% of cases showing partial divergence for \SE-on-\PSY{} and \PSY-on-\SE{}, respectively.
	
	\GPT's performance in consistency exceeded our expectations, showing less than 10\% disagreement across multiple assessments. This high degree of internal consistency may be due to the isolated nature of the tasks we assigned, \ie, tasks that did not require a complex reasoning or a deep understanding of multifaceted research. However, the underlying causes behind \GPT's consistency are worth exploring in future research.

	The consistency levels among human reviewers reinforce our observation that consistently evaluating the relevance of diverse papers can be a challenging task, particularly in cases of extensive, interdisciplinary research. It is important to note that the observed disagreement rates among human reviewers stem from a relatively small sample size, which contrasts with the much larger number of papers identified in the original Scopus queries. Here, too, the factors driving these disagreements will be the subject of further study.
	
	\answerblock{
	\header{In answering \RQ2,} \GPT{} demonstrated impressive internal consistency (over 90\% self-agreement) and was able to accelerate relevance screening significantly using only paper titles. While it tended to be more conservative than human reviewers, especially regarding user-focused topics, the resulting 13\% disagreement rate was deemed acceptable for complex interdisciplinary assessments. Overall, this underscores \GPT's usefulness in large-scale \scopingreview{}, making it actionable and efficient within a meaningful timeframe while also identifying promising themes and gaps for future research.
	}
	
	\section{Threats to Validity}
	
	A number of construct validity threats affect our study at a fundamental level. Note that, in the literature, there are papers devoted especially to the techniques to identify and report threats to validity, \eg, \cite{threats-to-validity,slr-process-ampat}. We follow \cite{threats-to-validity}.

	First, relying solely on paper titles risks misrepresenting the actual content of papers, as titles may not always reflect the full scope or focus of the work. To mitigate this, we manually validated a representative sample of \GPT-screened titles against independent human judgments to assess alignment and detect potential misclassifications. Terminological ambiguity also poses a challenge: key terms may carry different meanings across \SE{} and \PSY, which could complicate assessments of relevance. We addressed this by designing inclusion questions that explicitly spanned both disciplinary lenses and by grounding our thematic analysis in interdisciplinary reasoning.
	
	Further threats arise from how we defined and operationalized the intersection between \SE{} and \PSY. In particular, the distinction between \SE-on-\PSY{} and \PSY-on-\SE{} may not always be meaningful or could inadvertently separate what should be treated as a unified body of work. To mitigate this risk, we interpreted these intersections as complementary viewpoints and evaluated the results with caution to avoid unwarranted dichotomies. Additionally, our selection of keywords and venues, which is central to defining the scope of the review, may not fully represent either field. We addressed this through an iterative process involving expert review, careful documentation, and publication of the full keyword and venue lists to support transparency and reproducibility.
	
	Several procedural and reliability-related threats also warrant attention. Our reliance on \GPT{} introduces the possibility of inconsistencies or biases, especially due to prompt design and the influence of batching, where titles shown together may shape each other's evaluation. To mitigate this, we crafted concise and standardized prompts based on the Persona pattern and executed three independent \GPT{} runs to assess consistency. We also manually reviewed edge cases to ensure critical decisions were not made in isolation. While Scopus provided a practical way to gather large-scale data, its limitations, \eg , incomplete indexing, opaque ranking algorithms, and a result cap of 8,000, could affect the completeness and ordering of our dataset. We treated Scopus not as an exhaustive source but as a means of broad sampling and documented all filtering choices and assumptions.
	
	Interpretation-related threats include the risk of drawing flawed conclusions from observed patterns, \eg , such as mistaking structural disciplinary differences for methodological outcomes. For instance, the imbalance we observed in \SE's adoption of \PSY{} methods versus the reverse may reflect deeper institutional or epistemological trends rather than issues in our approach. We remained cautious in interpreting such findings, framing them as hypotheses for future inquiry rather than definitive conclusions. Additionally, our use of ratios, tag distributions, and other statistics are intended for descriptive purposes only; we avoided making inferential claims based on these figures.
	
	Finally, threats related to human judgment and validation must be acknowledged. Our manual review process involved a small number of reviewers and experts, all with \SE{} backgrounds and connected to the project. While this raises concerns about bias and limited disciplinary perspective, we addressed it by applying structured consolidation guidelines, resolving disagreements collaboratively, and documenting decisions consistently. Although the manual validation sample size is limited, it served as a grounded reference point for evaluating \GPT's performance and ensuring practical relevance.
	
	Overall, while these threats introduce uncertainty, we approached them with care through a combination of transparent methodology, multi-pass validation, and reflective interpretation. Our aim was not to eliminate ambiguity entirely but to make our assumptions explicit and ensure the process remains open to scrutiny and refinement.
		
	\section{Conclusion and Further Work}
	
	This interdisciplinary \scopingreview{} provided a broad, bird's-eye view of how verbalization is applied at the intersection of software engineering (\SE) and psychology (\PSY). Leveraging \LLM-assisted screening across over 9,000 papers, we efficiently assessed the relevance and clustered themes using only paper titles. \GPT's performance was statistically validated against human reviewers, showing high internal consistency and a manageable 13\% disagreement rate. The most prominent themes centered on expert cognition, decision-making, and collaboration, particularly in \SE{} literature. We observed a notable asymmetry, which we interpret as \SE{} frequently adopting \PSY{} methods, while \PSY{} rarely targets \SE-specific challenges. Moreover, the data suggests verbalization techniques are primarily used to capture behavior rather than to analyze or conceptualize it. Underrepresented themes (such as diversity, implicit human factors, and historical context) highlight research gaps and point to a broader trend: we assume \SE{} tends to apply human-centered methods pragmatically, whereas \PSY{} has less structural incentive to engage with \SE{}.
	
	The findings of this review carry several implications for \SE{} research. The dominant use of verbalization to capture behavior rather than to analyze or conceptualize it may limit the development of richer theoretical insights into human factors. The observed asymmetry between \SE{} and \PSY{} points to an opportunity for more balanced, reciprocal collaboration, \eg , if \SE{} researchers engage more deeply with psychological theory to strengthen methodological rigor. While the focus on practical themes aligns with \SE's outcome-driven culture, it risks neglecting subtler human dynamics, such as diversity, implicit biases, and historical context. Addressing these underrepresented areas could lead to more inclusive and reflective \SE{} practices. Finally, our use of \LLM-assisted screening demonstrates the potential of \AI{} tools to make large-scale, interdisciplinary scoping more feasible, laying the groundwork for more focused, follow-up reviews.
	
	As a next step, we aim to deepen our exploration of the most prominent themes identified in this review, synthesizing scattered knowledge into more cohesive frameworks that can inform both \SE{} theory and practice. At the same time, the underexplored themes deserve targeted attention to broaden the human-centered dimension of \SE{} research. Further methodological improvements are also in scope: by analyzing \GPT's decision patterns and failure cases, we hope to refine \LLM-assisted review pipelines and establish practical heuristics for conducting future large-scale interdisciplinary reviews.

	These efforts converge toward our broader goal: developing a structured catalog of verbalization-based techniques for \SE{} research. This catalog will bridge \PSY{} \enquote{know-how} and theoretical grounding with \SE{} \enquote{utilization} and open research challenges. It will organize exemplary studies from both fields, provide technique summaries, and link them to unresolved \SE{} questions where verbalization could offer insight, especially around cognition, decision-making, and behavior. While not exhaustive, the catalog is intended as a practical resource to help \SE{} researchers adopt and adapt verbalization methods meaningfully. Collaboration with \PSY{} experts will be a key to enhancing its validity and depth, ensuring it remains both accessible and rigorous.
	
	\begin{credits}
	\subsubsection{\ackname}
	This research was supported by the University Researcher Scholarship Program of the Ministry of Culture and Innovation, funded by the National Research, Development, and Innovation Fund (Grant ID: EKÖP-24-4-SZTE-607).

	\subsubsection{\discintname}
	The authors declare that they have no known competing financial interests or personal relationships that could have appeared to influence the work reported in this paper.

	\end{credits}
	
	\bibliographystyle{splncs04}
	\bibliography{main}

\end{document}